\documentclass[twocolumn]{aastex631}

\usepackage{amsmath}
\renewcommand{\vec}[1]{{\boldsymbol #1}}
\newcommand{\B}{\vec{B}}
\newcommand{\E}{\vec{E}}
\newcommand{\V}{\vec{v}}
\newcommand{\Vp}{v_{\rm p}}
\newcommand{\RL}{R_{\rm L}}
\newcommand{\RY}{R_{\rm Y}}
\newcommand{\BL}{B_{\rm L}}
\newcommand{\BY}{B_{\rm Y}}
\newcommand{\Bp}{B_{\rm p}}
\newcommand{\z}{\xi}
\newcommand{\exx}[2]{#1 \times 10^{#2}}
\shorttitle{centrifugal acceleration}
\shortauthors{Shibata \& Kisaka}

\begin{document}

\title{The centrifugal acceleration and the Y-point of the Pulsar Magnetosphere}

\author[0000-0002-8554-9240]{Shinpei Shibata}
\affiliation{
Faculty of Science, Yamagata University \\
Kojirakawa 1-4-12, Yamagata 990-8560, JAPAN}

\author[0000-0002-2498-1937]{Shota Kisaka}
\affiliation{
Physics Program, Graduate School of Advanced Science and Engineering, Hiroshima University,\\
Higashi-Hiroshima 739-8526, Japan}

\begin{abstract}
We investigate the centrifugal acceleration in an axisymmetric pulsar magnetosphere
under the ideal-MHD approximation.
We solved the field-aligned equations of motion
for flows inside the current sheet with finite thickness.
We find that flows coming into the vicinity of a Y-point become super fast.
The centrifugal acceleration takes place efficiently,
and most of the Poynting energy is converted into kinetic energy.
However, the super fast flow does not provide enough
centrifugal drift current to open the magnetic field.
Opening of the magnetic field is possible by the plasmas
that are accelerated in the azimuthal direction with a large Lorentz factor
in the closed field region.
We find that this acceleration takes place if the field strength increases
toward the Y-point from inside.
The accelerated plasma is transferred from the closed field region to the open field region
by magnetic reconnection with plasmoid emission. 
We also estimate the Lorentz factor to be reached in the centrifugal wind.
\end{abstract}

\keywords{
Rotation powered pulsars (1408) ---
High energy astrophysics (739) ---
Magnetohydrodynamics (1964) ---
Plasma astrophysics (1261)
}

\section{Introduction} \label{sec:intro}

Particle acceleration by pulsars is
a long-standing issue in astrophysics.
In an early stage of investigation,
the idea of the relativistic centrifugal wind is intensively studied
\citep{1969ApJ...158..727M,1970ApJ...160..971G,1994MNRAS.270..687L}.
Since the centrifugal acceleration is based on a corotational motion of the
magnetospheric plasmas, plasma density is assumed to be high, and
the ideal-MHD condition holds everywhere.
In reality, however, acceleration mechanism depends on plasma density,
and more specifically on where and how much electron-positron pairs are created.
Then particle acceleration by field-aligned electric field also interests us 
to explain the pulsed emission from radio to gamma-ray 
\citep{1976ApJ...206..831J,
1976ApJ...203..209C,
2011MNRAS.415.1827T,
1975ApJ...196...51R,
1997MNRAS.287..262S,
2010MNRAS.408.2092T}.
Recently, Particle-in-Cell (PIC) simulations are applied to understand 
the global structure of the pulsar magnetosphere 
\citep{2014ApJ...795L..22C, 2015MNRAS.449.2759B, 2023ApJ...943..105H, 2023ApJ...958L...9B, 2022ApJ...939...42H}.
They are able to treat both the field-aligned particle acceleration
and the centrifugal acceleration.
However, the mechanism of the centrifugal acceleration still remains unresolved.
In this paper, 
we revisit this problem, 
and compare with recent PIC simulations.

The model we consider is an axisymmetric steady magnetosphere with
a magnetic dipole moment at the center parallel to the angular velocity.
The plasma number density is assumed to be much higher than the Goldreich-Julian
density everywhere, so that the ideal-MHD approximation holds.
We do not deal with pair creation processes.

The magnetospheric plasma tends to corotate with the star.
Strict corotation would lead us to a divergence of the Lorentz factor
at the light cylinder, i.e.,
\begin{equation}
v_\varphi = \varpi \Omega_* \rightarrow c 
\ \ \ \mbox{as }
\varpi \rightarrow c/\Omega_* \equiv \RL ,
\end{equation}
where $\Omega_*$ is the angular velocity of the star,
$\varpi$ is the axial distance, and $R_L$ is the radius of the light cylinder.
In reality, just within the light cylinder,
the inertia would be  so large that
the centrifugal drift current would open the magnetic field lines.
The plasma accelerated in azimuthal direction with a large Lorentz factor
would finally be thrown away along the open field lines.
This is the idea of the centrifugal acceleration,
just like a trebuchet.

\begin{figure}
\includegraphics[width=\columnwidth, page=1]{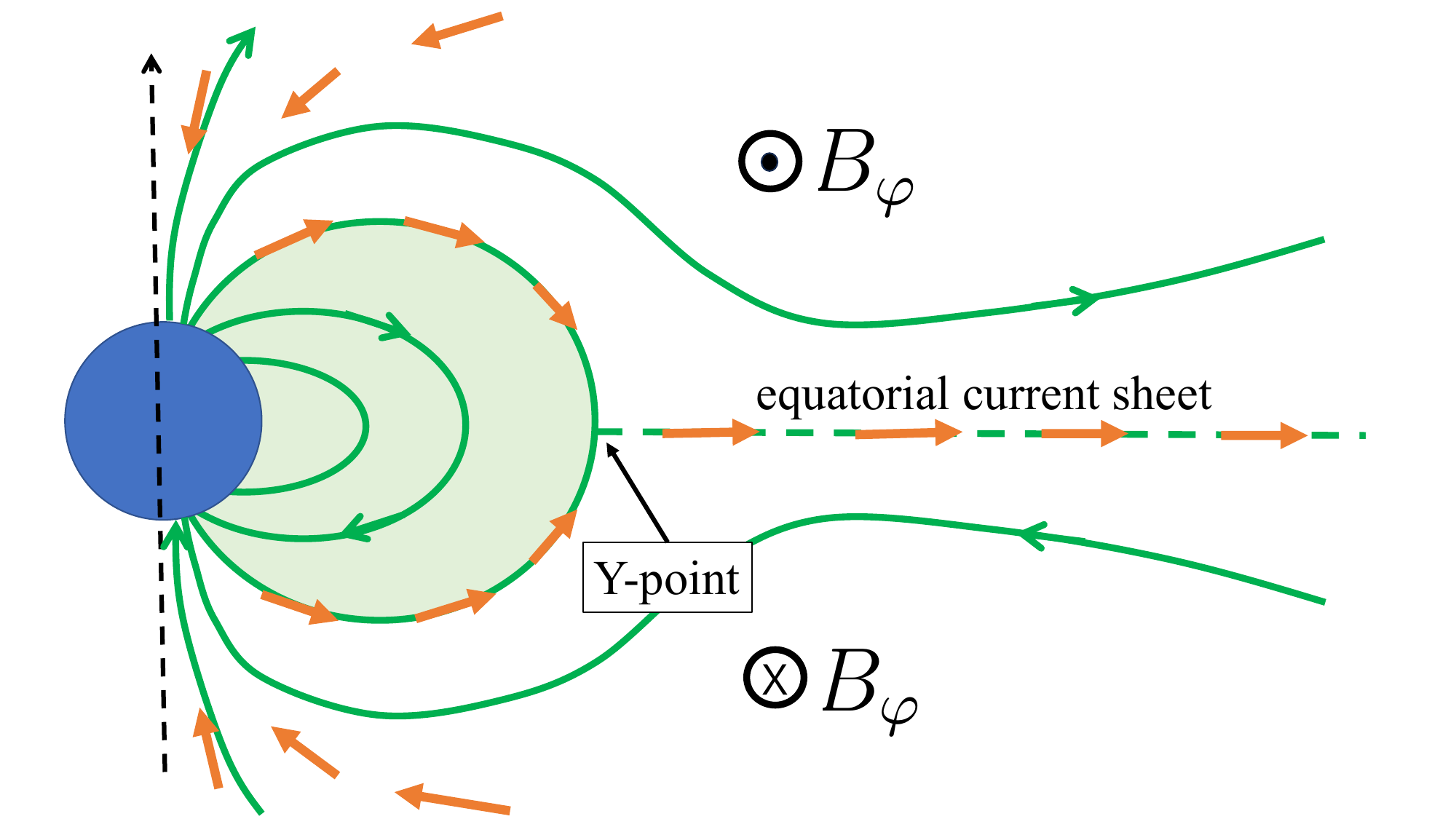}
\caption{Schematic picture of an axisymmetric magnetosphere 
with close and open magnetic field lines. 
The small arrows indicate poloidal currents, 
which form a loop in each hemisphere 
starting from the star in lower latitudes 
and go back to the star in higher latitudes. 
The top of the close field lines
is a junction of the current sheets 
in the force-free limit, called a Y-point.}
    \label{p-fig1}
\end{figure}

However, after some works
\citep[eg.,][]{1969ApJ...158..727M,1994ApJ...426..269B,1998PASJ...50..271T},
it is concluded that
the centrifugal acceleration is inefficient, and
flows on the open field lines are Poynting energy dominant.
There is  toroidal magnetic field $B_\varphi$
due to the global current loop 
as shown in Figure~\ref{p-fig1}.
The azimuthal velocity does not follow the corotation but
does the iso-rotation law, $v_\varphi = \Omega_* \varpi + \kappa B_\varphi$,
where $\kappa$ is a scalar function to be determined.
The open magnetic field lines trail backward due to rotation,
resulting in negative  $\kappa B_\varphi$, meaning departure from the corotation.
Thus the slingshot acceleration fails.
A typical Lorentz factor to be reached is 
$\sigma_*^{1/3}$ \citep[ and elsewhere]{1969ApJ...158..727M} 
where $\sigma_* = B^2 / 4 \pi nmc^2 $,
is the magnetization parameter,
and $B$ and $n$ are the field strength and the number density, respectively.
An expression $\sigma_* = \gamma_{\rm max}/2 {\cal M}$ may be useful,
where 
$\gamma_{\rm max} = e \BL /mc \Omega_*$ is the maximum reachable Lorentz factor,
${\cal M} = n /(\Omega_* \BL /2 \pi ec)$ is 
the multiplicity of the plasma, 
and $\BL$ is the light cylinder field. 
The value of $\sigma_*$ gives a typical Lorentz factor
when most of the Poynting energy is converted into plasma.
A flow with the Lorentz factor of $\sigma_*^{1/3}$ is obviously Poynting energy dominant.

The poloidal current loop has an essential role to carry Poynting energy away from the neutron star,
since it generates $B_\varphi$ and the Poynting flux,
$c \E_\perp \times \B_\varphi / 4 \pi $,
where $\E_\perp$ is the electric field across the magnetic field produced 
by the electromotive force of the star.
Due to symmetry, however, 
the equatorial plane is the place where the toroidal field vanishes.
In a region where the toroidal field becomes very weak,
the plasma motion may become very close to the pure corotation 
to get large Lorentz factors.
Such a place is where
the equatorial current sheet touches the closed field region,
a vicinity of a so-called  Y-point
(Figure~\ref{p-fig1}).

Meanwhile, highly accurate PIC simulations 
are available to provide detailed analysis for the centrifugal
acceleration 
\citep{2014ApJ...795L..22C, 2015MNRAS.449.2759B, 2023ApJ...943..105H, 2023ApJ...958L...9B, 2022ApJ...939...42H}.
It has been shown that
most of the space outside the equatorial current sheet is
Poynting energy dominant. However,
Poynting energy is converted to kinetic energy
in the vicinity of the Y-point and the equatorial current sheet.
It is interesting that
\citet{2022ApJ...939...42H} show 
an acceleration in azimuthal direction such as expected in the
centrifugal acceleration. Another interesting feature of their simulations
is ejection of plasmoids.

In this paper,
we study the centrifugal acceleration
that is expected in a vicinity of the Y-point.
It will be shown that super-fast flows appear inside the thin current layer.
The centrifugal drift current opens the poloidal magnetic fields in the
vicinity of the Y-point.
We will propose a self-consistent structure of the centrifugal wind.

\section{Magnetic field opening by the centrifugal drift current}

Let us estimate the Lorentz factor to open the magnetic field lines
by centrifugal drift currents.
The region we consider is a vicinity of the Y-point.
In the following sections,
we use the cylindrical coordinate $(z, \varpi, \varphi)$ with
the unit vectors denoted by 
$(\vec{e}_z ,  \vec{e}_\varpi ,  \vec{e}_\varphi )$.
The centrifugal drift velocity, which has opposite directions
depending on the charge sign, would be
\begin{equation} 
\V_{\rm d} = c \frac{\vec{F} \times \B}{\pm e B^2}
\approx \pm \frac{ \gamma mc^2}{e \BY \RY }
c \vec{e}_\varphi ,
\end{equation}
where the centrifugal force on an electron of mass $m$ and 
the Lorentz factor $\gamma$ is evaluated as 
$\vec{F} = (\gamma mc^2 /\RY ) \vec{e}_\varpi$,
$\RY \sim \RL$ is the axial distance of this region,
and 
$\BY$ is the magnetic field that is weakened significantly
as compared with $\BL$.
The drift current density may be
$j_\varphi = 2 e n |\V_{\rm d}| \approx
2nc (\gamma mc^2 / \BY \RY)$.
Applying Amp\`{e}re's law for a closed loop in the $\varpi - z$ plane 
around the Y-point,
we have $(4 \pi /c )j_\varphi \Delta = 2 \Bp^{\rm (out)} $
for a condition of opening the magnetic field, where
$\Bp^{\rm (out)} \approx \BL$ is the poloidal field strength outside the current layer, 
and $\Delta$ is a typical thickness.
With these conditions, we have the Lorentz factor 
to open the magnetic field,
\begin{equation} \label{eqdriftcurrent}
\gamma_c \approx \frac{(\Bp^{\rm (out)})^2}{4 \pi nmc^2} 
\frac{\RY}{\Delta} \frac{\BY}{\Bp^{\rm (out)}}
= \sigma_*
\frac{\RY}{\Delta} \frac{\BY}{\Bp^{\rm (out)}} .
\end{equation}

We expect a thin layer, $\Delta \ll  \RY$, and weak magnetic field,
$\BY \ll  \BL$. 
If $(\RY / \Delta )(\BY /\BL) \approx 1$,
$\gamma_{\rm c} \approx \sigma_*$. 
It is one of our objectives 
to find $\Delta$ and $\BY$ in a self-consistent manner, 
considering the magnetohydrodynamics inside the current layer.

\section{Current distribution in the current sheet}

In the following calculations of field-aligned flows, 
we assume an axisymmetric magnetosphere with
the ideal-MHD condition.

\begin{figure}
\includegraphics[width=\columnwidth, page=2]{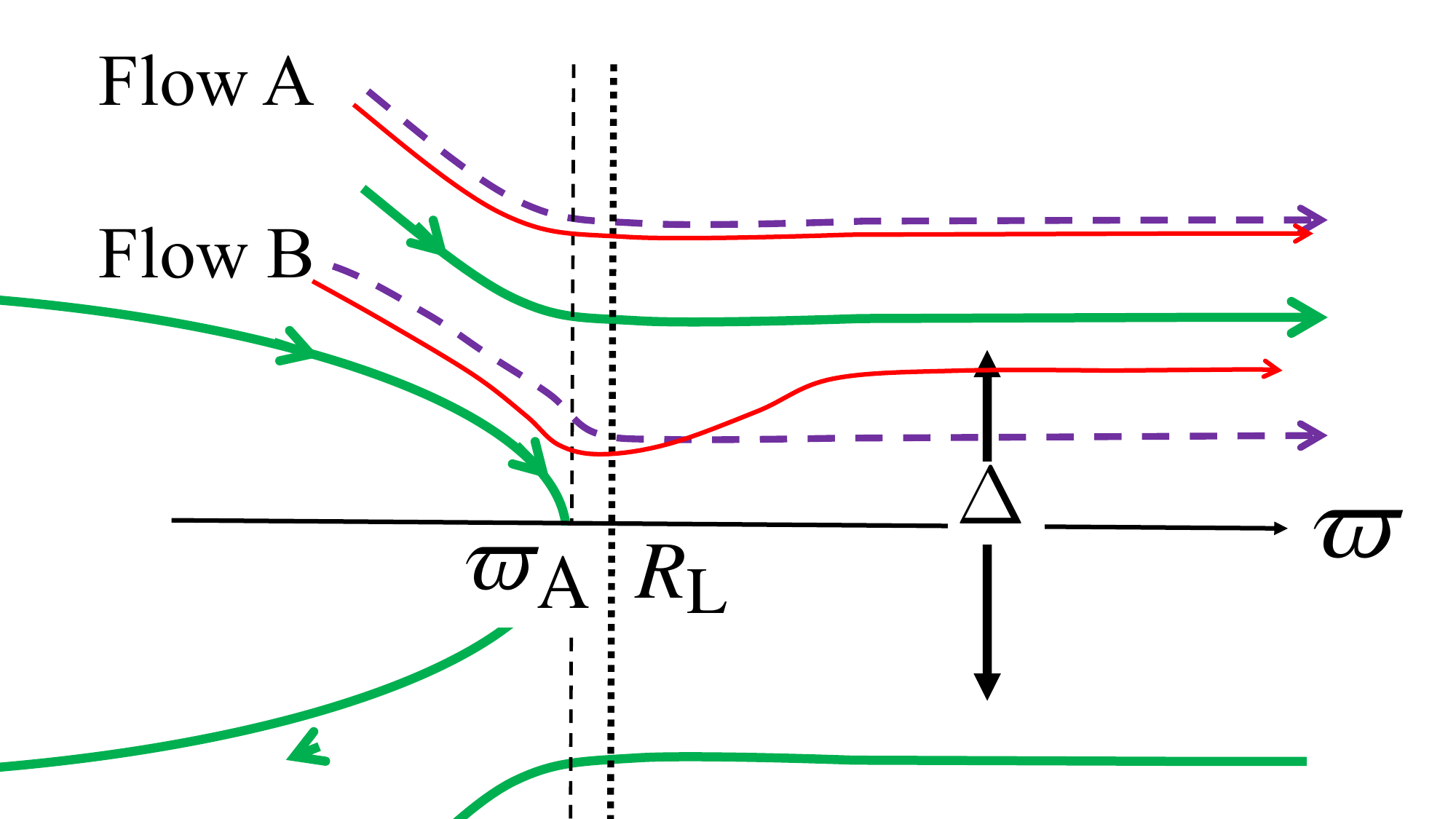}
    \caption{Two types of field-aligned flows. 
Flow A and Flow B, respectively, runs outside and inside
the current layer with the thickness $\Delta$.
The dashed curves are their stream lines.
The solid curves are the magnetic field lines.
The thin red curves are the poloidal current lines.
The $\varpi$-axis indicates the axial distance.
The thin dashed line $\varpi = x_{\rm A}$ is the Alv\'{e}n surface.
The thin dotted line $\varpi=\RL$ is the light cylinder.
}
    \label{p-fig3}
\end{figure}

Let us consider two flows:
one runs just outside the equatorial current layer,
the other runs inside. 
We call them Flow A and Flow B, respectively,
as shown in Figure~\ref{p-fig3}.
{What is distinctive of Flow B from Flow A is
significant decrease in the poloidal field strength 
as is flows near the Y-point. }
We expect that 
the centrifugal acceleration is inefficient in Flow A, 
but efficient in Flow B.

Flows such as Flow A and Flow B can be solved by using the field-aligned equations 
\citep{1978MNRAS.185...69O, 1986A&A...156..137C, 1991PASJ...43..569T},
provided that a poloidal field is given.
It must be noticed here that
the solution of field-aligned flows determines
values of $B_\varphi$ and in turn the current function $I=\varpi B_\varphi$,
where stream lines of the poloidal current is given by contours of $I$.
The field-aligned equations are solved with 
a given set of boundary conditions and the condition 
that the solution passes through the fast critical point.
As a result, $I$ is uniquely determined.
On the other hand, as was done in obtaining the force-free solution
\citep{1999ApJ...511..351C, 2006MNRAS.368.1055T},
the trans-field equation that gives the poloidal field structure
determines the current function $I$ so that
the field lines are regular on the Alfv\'{e}n surface,
which coincides with the light cylinder in the force-free limit.
Thus, the both systems of equations can determine $I$.
The pulsar problem is only solved simultaneously 
the field-aligned equations and the trans-field equation, i.e.,
the two $I$'s obtained in the two ways must be identical.
This has never been done so far.

The force-free solution may be a good approximation 
for the field structure in
high and middle latitudes, because 
the field-aligned solutions give
weak centrifugal acceleration there.
This is supported by the PIC simulations and
the Relativistic Magnetohydrodynamic simulations, which give the results
very similar to the force-free solution.

The problem is how the current runs
inside the current layer and in the vicinity of the Y-point.

For flows such as Flow A,
the poloidal current lines almost coincide with the flow stream lines,
since the centrifugal acceleration is weak.
If a strong acceleration takes place on Flow B,
the current stream lines crosses the flow stream line causing the acceleration.
In this case,
some current lines converge to the Y-point as indicated in 
Figure~\ref{p-fig3}, 
forming a very thin current sheet.
This seems a realization of 
the infinitely thin current sheet that is required by the
regularity on the light cylinder in the force-free approximation.
Even if inertia is taken into account in the trans-field equation,
the inertial effect may not be significant on the Alfv\'{e}n surface;
the acceleration become significant beyond the Alfv\'{e}n surface.
Therefore, the thin current sheet on the Alfv\'{e}n surface would be  
a genuine property of the centrifugal wind, ensuring the regularity
of the poloidal magnetic field.

\section{Field-aligned flows in the current layer}

In this section, we solve the field-aligned equations
for Flow A and B.
The field-aligned equations are composed of the three conservation laws
and the iso-rotation law.
Given a stream function $\psi$
of the poloidal magnetic field, 
the particle flux $g(\psi)$, the angular momentum $\ell (\psi)$, 
and the energy $\epsilon (\psi)$ are functions of $\psi$ only.
In the iso-rotation law, the angular velocity $\Omega (\psi)$ is also a function of $\psi$.
These quantities are all determined by a set of boundary conditions and the
condition that the flow passes through a fast-critical point.

The field-aligned equations are 
algebraic equations, and are joined into a single expression called
the Bernoulli function $\epsilon (\z,v)$.
Here, we normalize the axial distance $\varpi$ and 
the poloidal 4-velocity $u_{\rm p}$ by the values 
at the Alfv\'{en} point, i.e.,
$\z = \varpi /\varpi_{\rm A}$ and  $v= u_{\rm p} / u_{\rm A}$.

{ 
The field-aligned equations are 
algebraic equations, and are joined into a single expression called
the Bernoulli function $\epsilon (\z,v)$.
Here, we normalize the axial distance $\varpi$ and 
the poloidal 4-velocity $u_{\rm p}$ by the values 
at the Alfv\'{en} point, i.e.,
$\z = \varpi /\varpi_{\rm A}$ and  $v= u_{\rm p} / u_{\rm A}$.
A non-dimensional form of the Bernoulli function $\epsilon (\z,v)$ with
the definition of the Alfv\'{e}n point 
\citep{1978MNRAS.185...69O, 1986A&A...156..137C, 1991PASJ...43..569T}
is given below, and
it is notable that 
a solution is obtained as a curve $\epsilon (\z , v)=$constant on
the $\z$-$v$ plane.
The Bernoulli function $\epsilon (\z,v)$ has
two non-dimensional parameters, 
\begin{equation}
\lambda = \frac{\Omega (\psi) \ell(\psi) }{\epsilon (\psi)}, 
\end{equation}
and
\begin{equation}
\sigma = \frac{\Omega (\psi)^2 B_{\rm p} \varpi^2 }{4 \pi mc^3 g(\psi)},
\end{equation}
and it has an singularity at the Alfv\'{e}n point, 
$\varpi_{\rm A} = \sqrt{\lambda}  c/\Omega(\psi) $.
The 4-velocity at the Alfv\'{e}n point becomes 
$u_{\rm A} = (1-\lambda) \sigma_{\rm A} /\lambda$, where
$\sigma_{\rm A}$ is the value of $\sigma$ at the Alfv\'{e}n point.
The Bernoulli function has a form
\begin{equation} \label{bern}
\left( \frac{\epsilon (\z,v)}{mc^2} \right)^2 
= {\frac{ (1 + u_{\rm p}^2 )}{(1 - \lambda  )^2 }}
\ \
{
\frac{[ 1 - \lambda \z^2 - (1-\lambda )v\z^2 ]^2 }{(1-v\z^2 )^2 - \lambda \z^2 (1-v)^2  } 
} ,
\end{equation}
where
\begin{equation}
u_{\rm p} = \frac{1- \lambda }{ \lambda } \sigma v . 
\end{equation}
A family of solutions is easily drawn as contours of $\epsilon (\z,v)$
for a given set of $\lambda$ and $\sigma$. 
}

We chose one solution among them that passes through the fast-critical point;
hereafter we denote the fast point as $(\z_{\rm f}, v_{\rm f})$.
At the same time, the flow energy $\epsilon_{\rm f} = \epsilon(\z_{\rm f}, v_{\rm f})$
is obtained as if it is an eigenvalue.

The Bernoulli function $\epsilon (\z,v)$ has the two parameters,
$\lambda$ and $\sigma$.
However, 
the value of $\lambda$ is determined implicitly 
by given { inner} boundary conditions.
Firstly we assume $\lambda$, and obtain a critical solution.
Then we examine whether the solution satisfies the set of given boundary conditions.
If not, we change $\lambda$ until the boundary conditions are satisfied.
{ The inner boundary conditions are 
those for a plasma source region.
More specifically, for Flow A and Flow B, they may be 
such that the plasma is injected well within the light cylinder $\xi \ll 1$
with the Lorentz factor $\gamma \sim 1$.
}

Thus, $\sigma$ is the only parameter that determines the flow.
It is decomposed into two factors:
\begin{eqnarray}
\sigma &=&  \sigma_0 \hat{B} , \\
\label{sigmazero}
\sigma_0 &=& 
\frac{ \BL }{4 \pi mc g(\psi)}  
= \gamma_{\rm max} 
   \left(   \frac{\Vp /c }{{\cal M}}        \right)_{\rm in}
   \left( \frac{\Omega_* }{\Omega (\psi) }  \right) 
\approx 
    \frac{\gamma_{\rm max} }{{\cal M}_{\rm in}}, \\
\hat{B} &=& \frac{ \Bp \varpi^2 }{ \BL \RL^2 },
\end{eqnarray}
where $g(\psi ) = n \kappa = n v_{\rm p}/B_{\rm p}$,
and 
${\cal M}_{\rm in}$ is the parameter of the plasma number density
at the { inner boundary}, 
i.e., 
${\cal M}_{\rm in} =n/[\Omega (\psi ) \Bp / 2 \pi c e ]$, 
{ evaluated at the inner boundary.} 
The configuration of the poloidal magnetic field
only comes into the parameter $\hat{B} \propto B_p \varpi^2$.

\begin{figure}
\includegraphics[width=\columnwidth, page=3]{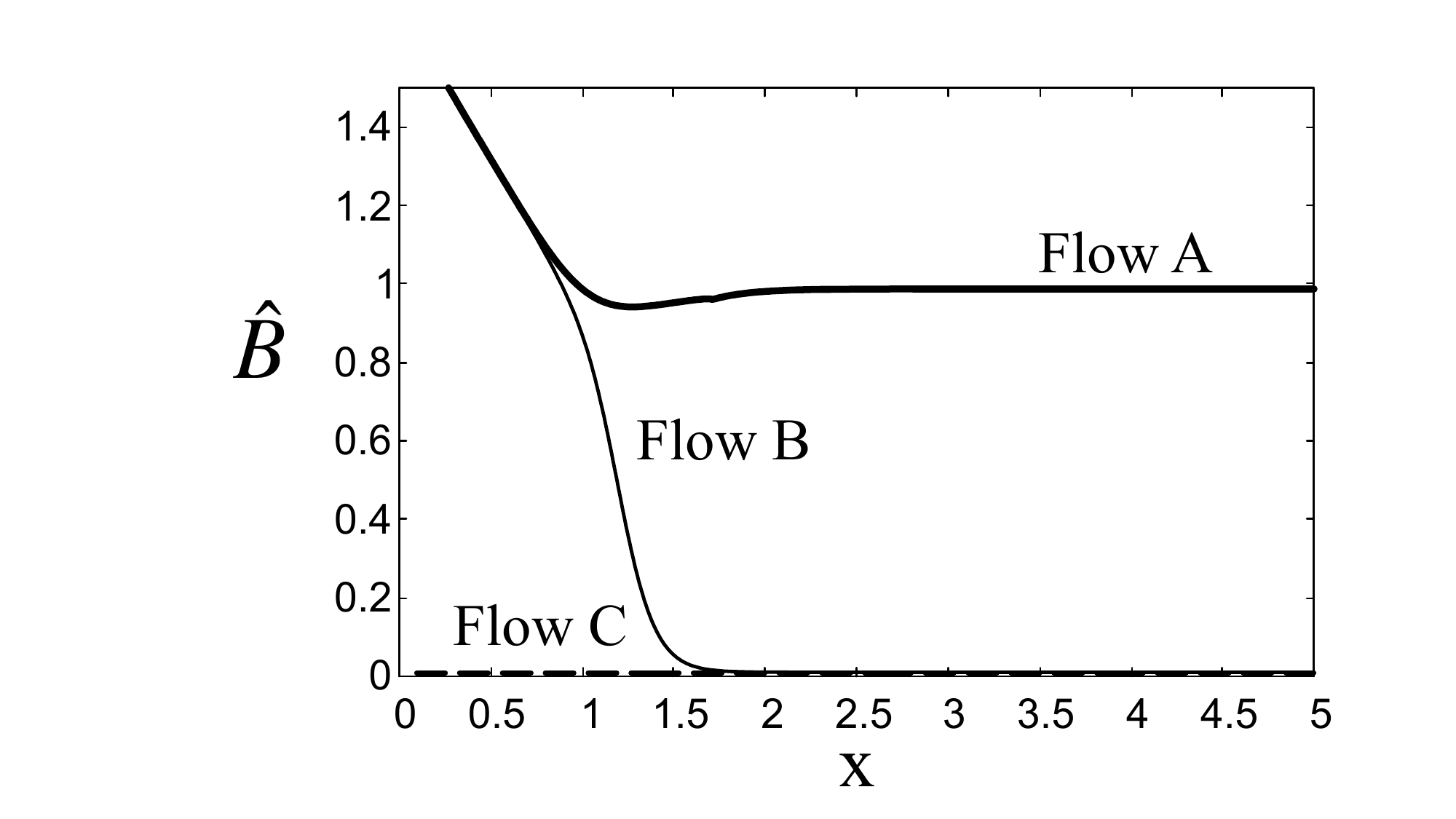}
    \caption{Models of $\hat{B}(x)  \propto B_{\rm p} \varpi^2$
for Flow A, Flow B and Flow C.}
    \label{p-fig5}
\end{figure}

The key parameter that makes 
difference between Flow A and Flow B
is $\hat{B}$.
The poloidal magnetic field would be something like dipole
near the star,
and changes to something like radial near and beyond the light cylinder.
Mimicking this, a simple function for $\hat{B}$ is introduced as follows.
The magnetic stream function for a dipole field
is $\psi = \BL \RL^3 \varpi /r^2$, where
$r= (z^2 + \varpi^2 ) ^{1/2}$.
For the dipole field, $\hat{B}$ is given by
\begin{equation}
\hat{B}_{\rm dip}  (x)
= ({\psi / \BL \RL^2}) 
\left[ 4 - 3 ({\psi x / \BL \RL^2} )^{2/3} \right]^{1/2} ,
\end{equation}
where $x= \varpi / \RL$.
Since $\hat{B}$ is constant for a radial field,
we assume that it gradually changes to a constant, 
$\hat{B}_{\rm radial}= \hat{B}_{\rm dip} (x_{\rm c})$ 
beyond a certain point $x_{\rm c}$. 
Then we assume that $\hat{B}$ for Flow A is 
\begin{equation}
\hat{B}_{\rm a} (x) 
= S \left(\frac{x_c-x }{\delta_c} \right) \hat{B}_{\rm dip} (x) 
+ S \left(\frac{ x-x_c }{ \delta_c} \right)  \hat{B}_{\rm radial},
\end{equation}
where $S (x) = 1/[\exp (-x) + 1 ]$ is the sigmoid function
(see Figure~\ref{p-fig5}).

\begin{figure}
\includegraphics[width=\columnwidth, page=4]{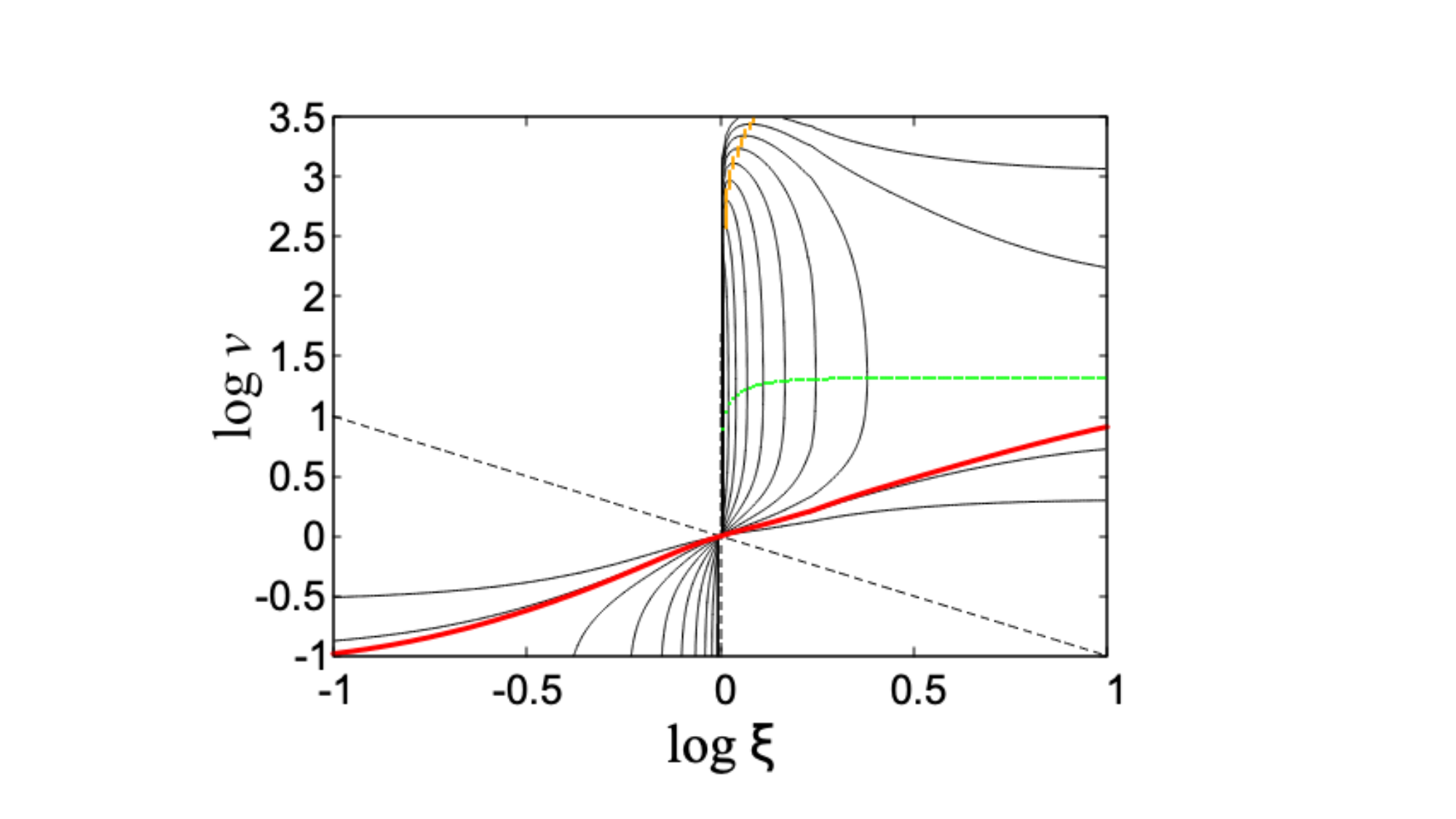} 
\caption{A family of the solutions and the critical solution (red)
of Flow A.
The dashed curves indicate the boundaries of the forbidden region.
The green and orange dots indicate where 
$\partial \epsilon / \partial v =0$
and 
$\partial \epsilon / \partial \z =0$, respectively.  }
    \label{p-fig6}
\end{figure}

As an example of Flow A, we take 
$\gamma_{\rm max} = 10^7$ and ${\cal M}_{\rm in} = 10^3$, and therefore
$\sigma_0 = 10^4$.
The solutions in the $\z$-$v$ plane are shown in Figure~\ref{p-fig6},
where the critical solution is colored in red.
It is { known}  that all the solutions pass through the Alfv\'{e}n point.
The denominator of (\ref{bern}) must be positive, so that
there is a forbidden region, which is bounded by the two curves,
\begin{equation}
v_+ = \frac{1+\sqrt{\lambda} \z }{\z(\z+\sqrt{\lambda})}, \ \ \ \ \ 
v_- = \frac{1-\sqrt{\lambda} \z }{\z(\z-\sqrt{\lambda})},
\end{equation}
indicated by the dashed curves in the figure.
The critical points are the points where the two curves,
${\partial \epsilon }/{\partial v} =0$ and
${\partial \epsilon }/{\partial \z} =0$ meet.
These curves are also shown with the dots in Figure~\ref{p-fig6}.
Since $\hat{B}$ is essentially constant beyond the Alfv\'{e}n point,
the X-type critical point, namely the fast point, goes to infinity 
as in the case of the radial flow.
After an iteration,
the value of $\lambda$ is determined so that the injection point
locates at about $\z=0.1$.
The obtained parameters of the critical solution is given in Table~\ref{tab1}.

\begin{table*}
\begin{center}
\caption{ \label{tab1} Some parameters for the critical solution.}
\begin{tabular}{lrrrrrrr} \hline
Type   & $\sigma_0$ & $\log \z_f$ & $\log v_f$ & $\epsilon_f$      & $1-\lambda$      & $1-x_{\rm A}$    & $\gamma$ \\
\hline
Flow A & $10^4$     & $\infty$   & $1.3275$   & $\exx{0.9905}{4}$ & $\exx{1.02}{-4}$ & $\exx{5.15}{-5}$ & $21.4 {\rm (a)}$  \\
Flow B & $10^4$     & $0.0525$   & $0.9075$   & $\exx{3.039}{3}$  & $\exx{4.0}{-4}$  & $\exx{2.0}{-4}$  & ${\exx{2.95}{3}} {\rm (b)}$ \\
Flow C (c) & $10^4$     & $1$        & $1$        & $10^2$            & $\exx{4.0}{-4}$  & $\exx{2.0}{-4}$  & $10^2$ \\
\hline
\multicolumn{8}{l}{(a): evaluated at $\z=100$, (b): evaluated at $\z=3.14$, (c):$\lambda$ is assumed.} 
\end{tabular}
\end{center}
\end{table*}

\begin{figure}
\includegraphics[width=\columnwidth, page=5]{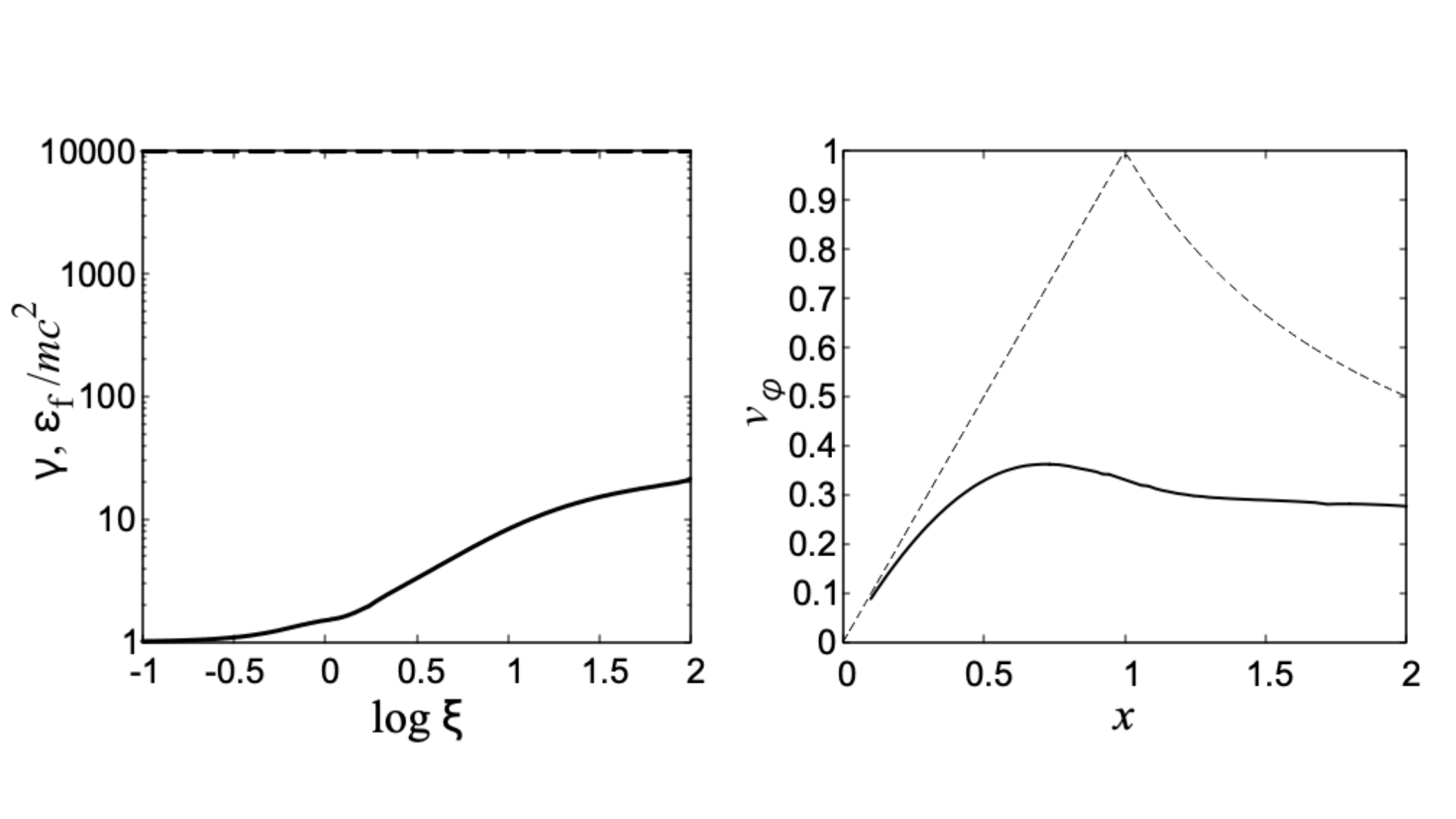}
    \caption{
Properties of Flow A.
Left panel: The particle energy $\gamma$ (solid curve), 
and the total energy $\epsilon_{\rm f}$ (dashed line)
as functions of $\log \z$.
Right panel: The azimuthal velocity $v_\varphi$ 
as a function of $x=\varpi / \RL$. 
The dashed curve indicates the corotation for $x <1$ and
the free emission tangent to the light cylinder for $x>1$.}
    \label{p-fig7}
\end{figure}

The critical solution gives $\gamma = 21.4$ at $\z=100$, 
which is consistent with the Michel's solution 
$\gamma \rightarrow \epsilon_{\rm f}^{1/3}= 21.5$ as 
$\xi \rightarrow \infty$ for the radial flow.
The total energy $\epsilon_{\rm f}$ is $\approx \sigma_0$.
In the left panel of Figure~\ref{p-fig7}, 
we plot $\gamma$ as a function of $\log \xi$ with
$\epsilon_{\rm f} /mc^2$, which is constant.
Flow A is a Poynting dominant flow.
The azimuthal velocity deviates from the corotation
as seen in the right panel of Figure~\ref{p-fig7}.

Next, we consider Flow B that
comes into a weak field region near the Y-point.
We expect a significant decrease of $\hat{B}$ beyond the Alfv\'{e}n point.
We model $\hat{B}$ by
\begin{equation} \label{Bhatmodel}
\hat{B}_{\rm b} (x) = 
\left[ (1-D ) \ S \left( \frac{x_{\rm d} - x }{\delta_{\rm d} } \right) + D \right]
\hat{B}_{\rm a},  
\end{equation}
where $D$ and $x_{\rm d}$, respectively, determine how much 
and where the magnetic field decreases.
If $D=1$, there is no decrease, but if $D<1$,
$\hat{B}$ drops down to $D \hat{B}_{\rm a}$ beyond $x_{\rm d}$.
In the following example, we use 
$D=0.01$, $x_{\rm d} = 1.2$, and $\delta_d = 0.1$, 
as shown in Figure~\ref{p-fig5}.
The value of $\sigma_0$ is the same as Flow A, i.e.,
$\sigma_0=10^4$.

\begin{figure}
\includegraphics[width=\columnwidth, page=6]{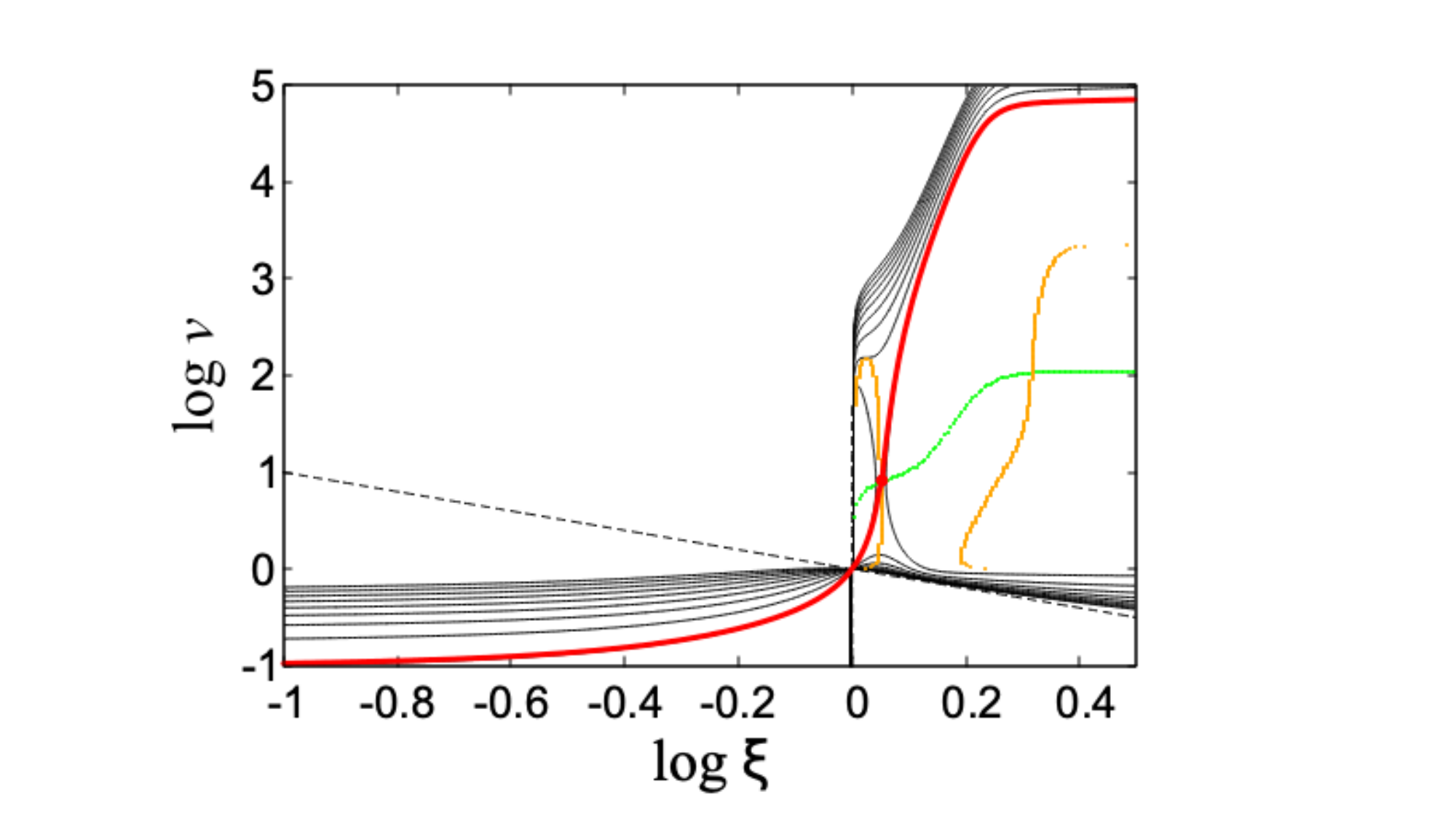} \\
    \caption{The same as Figure~\ref{p-fig6}, but for the flow B.}
    \label{p-fig8}
\end{figure}

\begin{figure}
\includegraphics[width=\columnwidth, page=7]{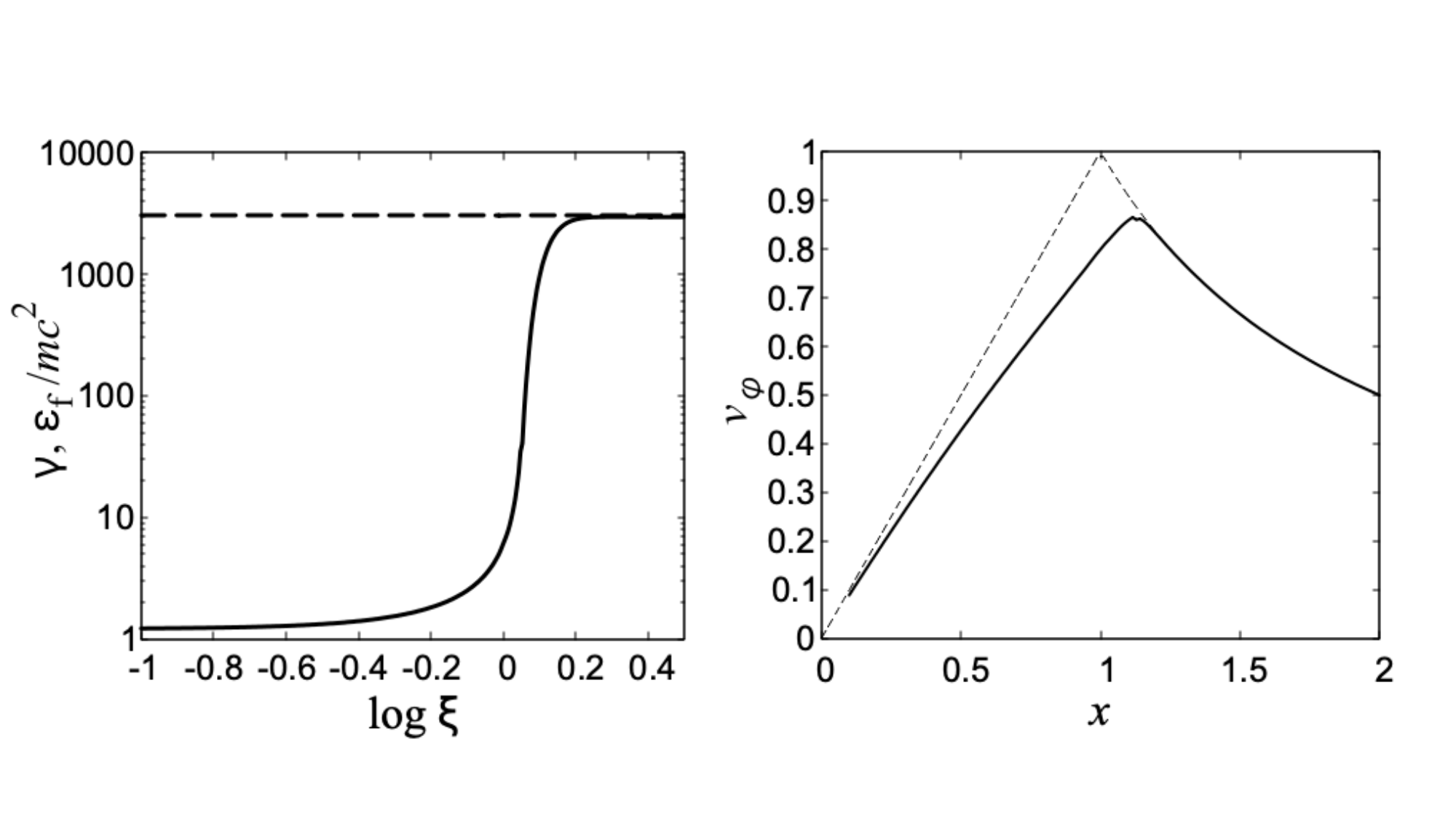} \\
    \caption{The same as Figure~\ref{p-fig7}, but for Flow B.}
    \label{p-fig9}
\end{figure}

There is a topological change in the solution curves 
as shown in Figure~\ref{p-fig8}. 
An X-type critical point, the fast point,  appears just beyond the Alfv\'{e}n point;
$\z_{\rm f} = 1.13$.
The Lorentz factor glows up around the fast point from the value at the injection
($\gamma =1.1$) to $\exx{2.95}{3}$, 
which corresponds to 98\% of the total energy $\epsilon_{\rm f}$ 
(Figure~\ref{p-fig9}).
In contrast with Flow A,
deviation of the azimuthal velocity from the corotation is weaker,
and $v_\varphi$ peaks at $\sim 0.85c$.
More importantly the curve eventually almost follows the analytic 
curve of the tangential ejection with the light speed at the Alfv\'{e}n radius.
The flow is effectively corotation up to the Alfv\'{e}n point.
The total energy $\epsilon$ is about 30\% of Flow A.
This is reasonable because 
the energy flux of Flow B is on its way of decreasing down to
zero on the equator.

It is known that
a decrease of $\hat{B}$ causes efficient conversion 
from the Poynting energy to kinetic energy
in the context of the AGN jets that have a collimated geometry
\citep{1989ASSL..156..129C, 1994ApJ...426..269B}.

\section{Centrifugal drift current of Flow B}

Our next question is whether the super fast flow such as Flow B can provide
enough centrifugal drift current to open magnetic field lies.
We take the same way as we derive (\ref{eqdriftcurrent}),
but we use
$v_{\rm d} = \pm (c/e) (\gamma m v_{\varphi}^2 / \RY \BY)$ 
and $n \Vp /\Bp  = g(\psi)$ to evaluate the density. 
Then we have
\begin{equation} \label{thickness}
\frac{\Delta}{\RY} =
\frac{(\Vp/c) }{ (v_\varphi/c)^2 } \ 
\frac{\sigma_0 }{\gamma } \ 
\frac{\BY}{\Bp} \ 
\frac{\Bp^{\rm (out)}}{\BL} .
\end{equation}
If the drift current is enough large, then
$\Delta / R_Y $ must be much smaller than unity.
However, this expression indicates that $\Delta / R_Y $ 
will never be small.
The last factor 
$\Bp^{\rm (out)}/{\BL}$ is about unity.
The factor ${\BY}/{\Bp}$ will not be much smaller than unity.
The effective field to produce the drift current is 
$z$-component of the poloidal field.
The region with a large azimuthal velocity would be
where the flow comes into the Y-point, and 
the ratio of the $z$-component to the poloidal component of the flow
will not be so small.
Moreover, if the Y-point is actually T-point as { suggested} by \citet{2024MNRAS.527L.127C},
it becomes unity.
The ratio $\sigma_0 / \gamma$ can be small, if 
$\gamma \gg \sigma_0$. 
However, a very efficient acceleration gives at most 
$\gamma \sim  \sigma_0$. 
Since $v_\varphi$ is something around $\sim 0.85$, 
the factor $(v_{\rm p}/c)/(v_\varphi /c)^2$ will
not be very small.
In conclusion, accelerated flows such as Flow B
will not produce enough current to
open the magnetic field.
This may sound strange when one compares 
with (\ref{eqdriftcurrent}).
However, as is seen in the expression
$n = g(\psi) \Bp / \Vp$, the plasma density decreases
due to the decrease of the magnetic field or increase of the cross section
of the flow.
As a result, even if the centrifugal acceleration becomes efficient,
the drift current density does not increase.

\section{Plasma injection at the Y-point}

\begin{figure}
   \includegraphics[width=\columnwidth, page=8]{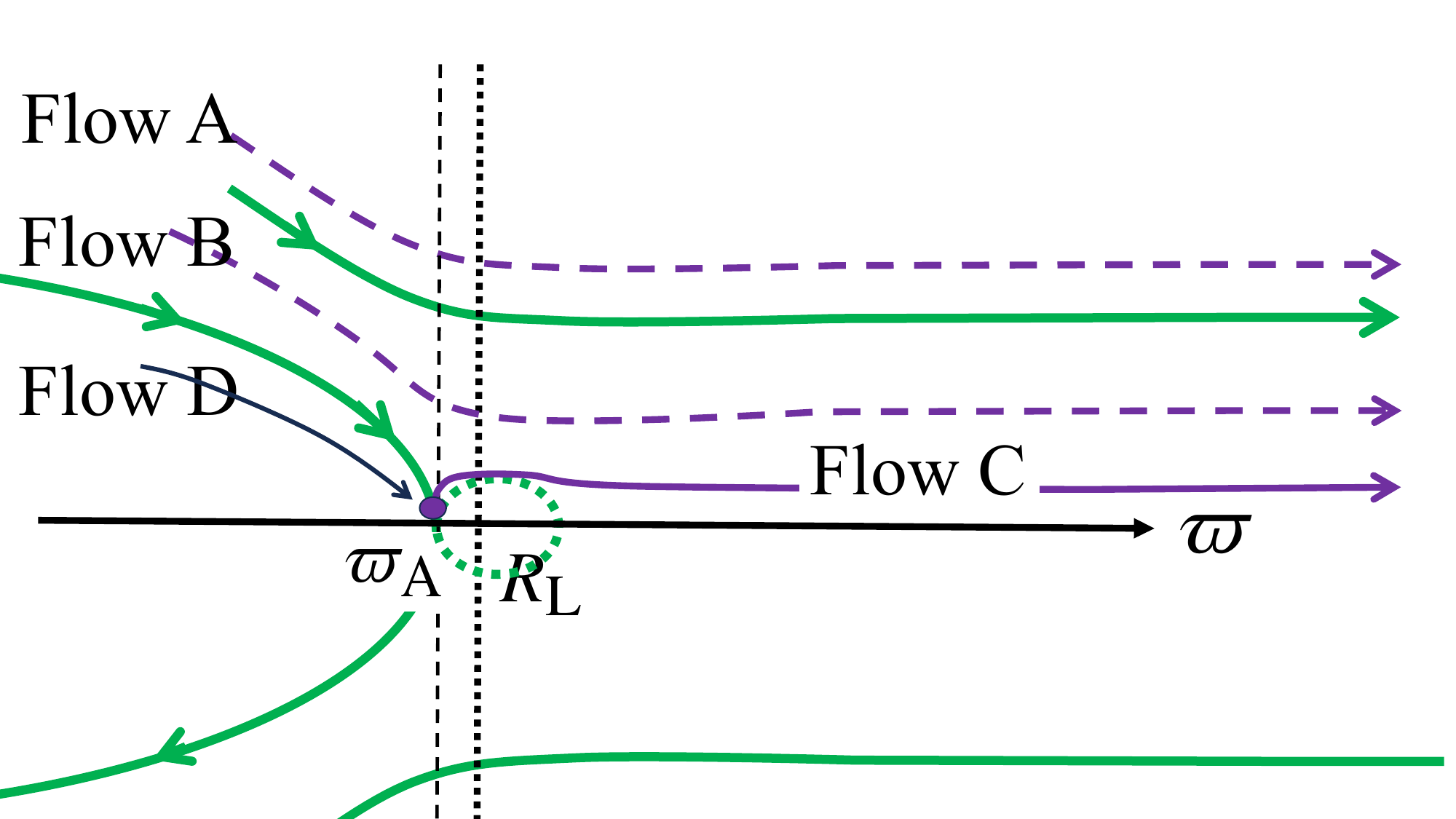}
    \caption{Flow C injected just within the Alfv\'{e}n point, and 
Flow D is a flow before the injection.}
  \label{p-fig10}
\end{figure}

The reason why sufficient drift current to open magnetic field lines
cannot be obtained is due to decrease in flow density
as the flow comes into the weak field region around the Y-point.
The magnetic field should be opened in the centrifugal wind.
One possible and very likely way to avoid the density decrease
is an injection of plasma at the Y-point.
Let us consider a flow injected from a top of the closed field region
into the current layer,
such as Flow C in Figure~\ref{p-fig10}.

The parameters characterize Flow C would be such that
$\hat{B}$ stays small throughout from injection (Figure~\ref{p-fig5}), 
and that $\sigma_0$ is also small.
In the definition (\ref{sigmazero}) of $\sigma_0$,
the density parameter ${\cal M}_{\rm in}$ is calculated at the injection
point where the magnetic field is weak around the Y-point.
Therefore,
${\cal M}_{\rm in}$ is much larger than a multiplicity based on
a typical Goldrech-Julian density, $\Omega_* \BL /2 \pi ce$.

For an example of Flow C,
we suppose that the magnetic field strength at the injection point is 
one hundredth of the field outside the current layer.
Then, 
$\sigma_0 = 10^4$ of Flow A and B is reduced to
$\sigma_0 = 100$ for Flow C, and 
$\hat{B}= \hat{B}_{\rm c}=\mbox{constant}= 0.01$
for Flow C.

\begin{figure}
\includegraphics[width=\columnwidth, page=9]{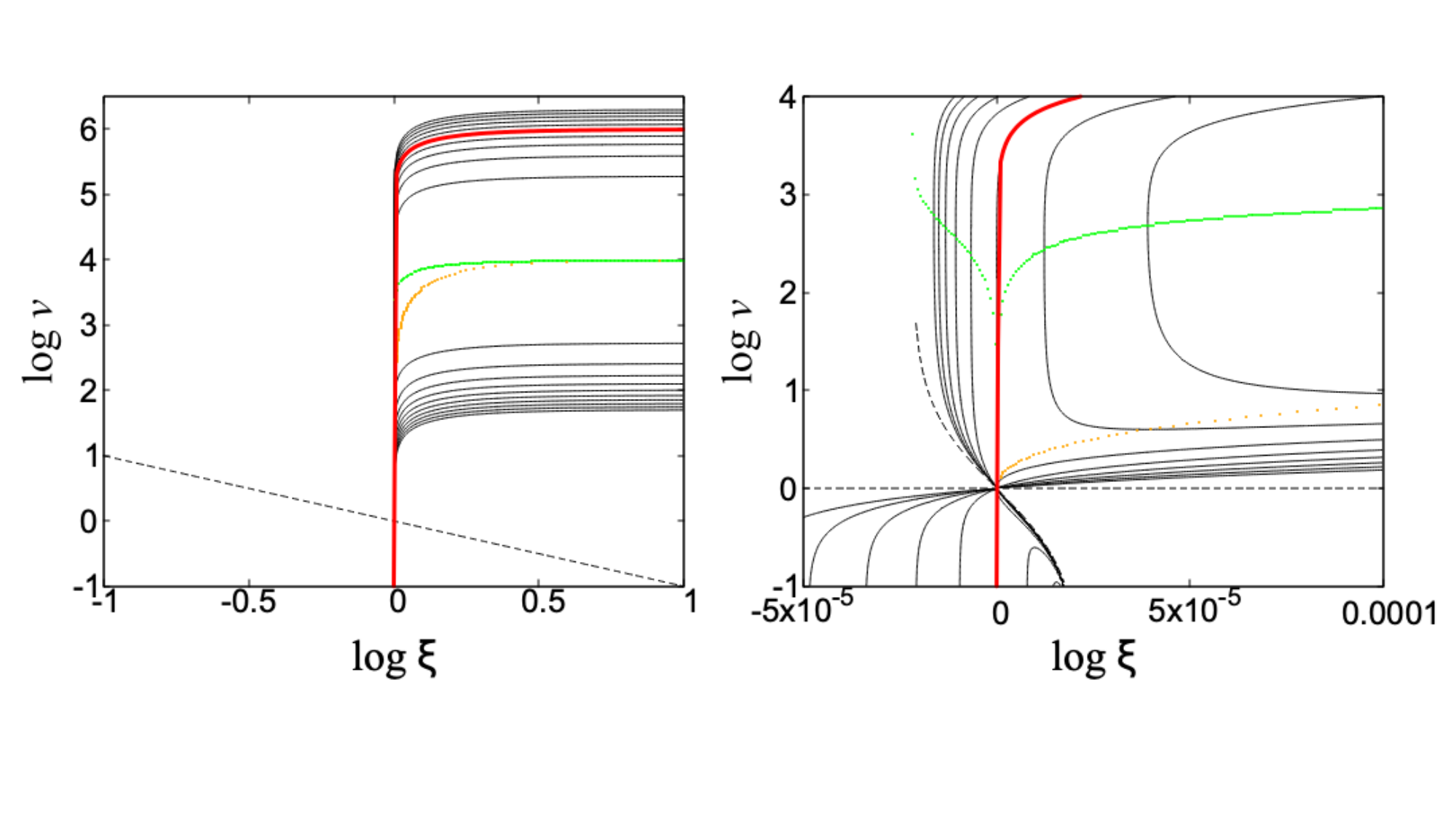}
    \caption{Left panel: The critical solution for Flow C injected 
just inside of the Alfv\'{e}n point.
Right panel: A close-up view around the Alfv\'{e}n point.}
    \label{p-fig11}
\end{figure}

\begin{figure}
\includegraphics[width=\columnwidth, page=10]{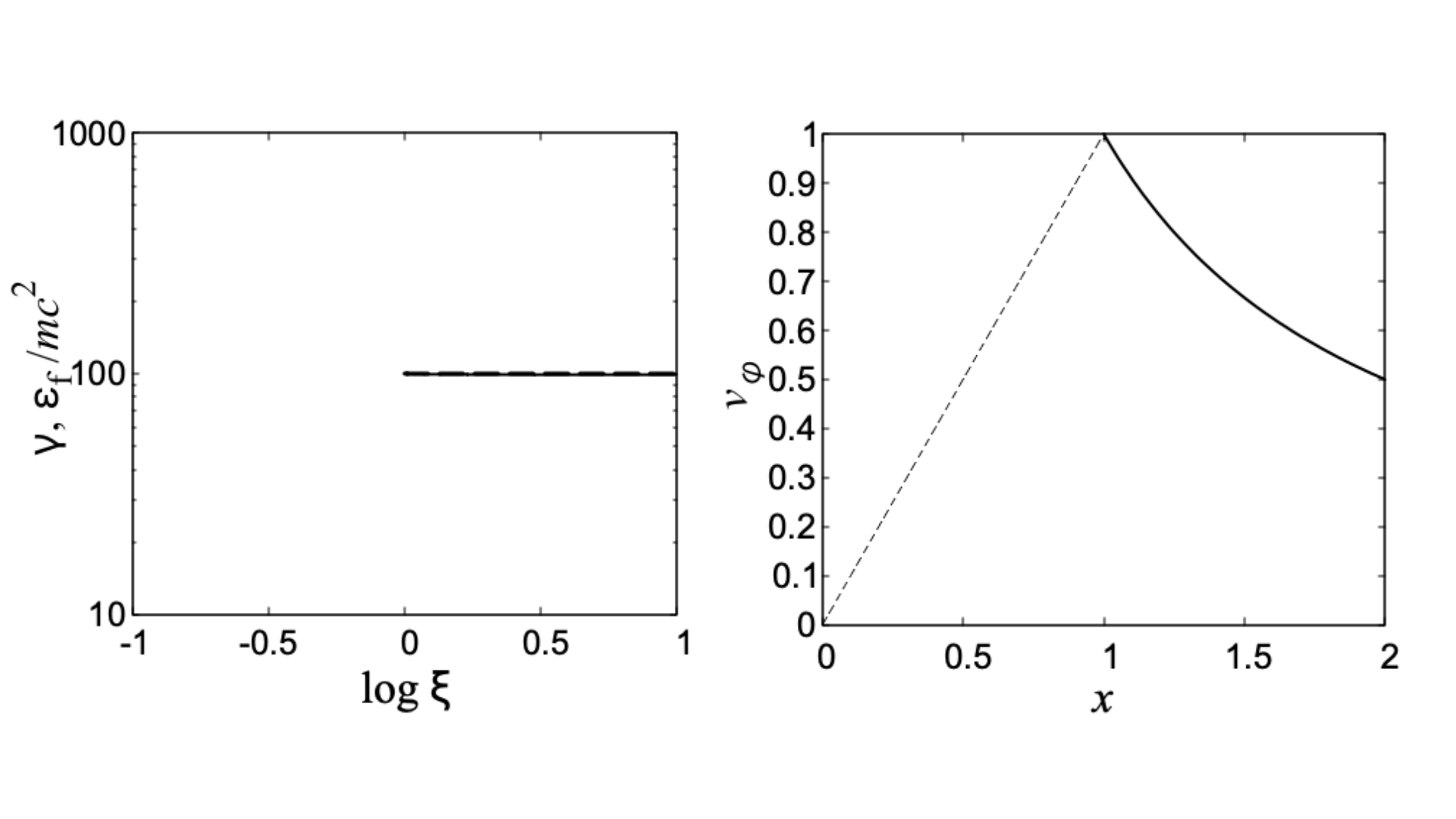}
    \caption{
Properties of Flow C.
Left panel: The Lorenz factor (solid line) 
and the energy of the flow (dashed line) as functions of $\log \z$.
Right panel: The azimuthal velocity as a function of $x$. }
    \label{p-fig12}
\end{figure}

A family of the solutions is shown in Figure~\ref{p-fig11}.
We find another topology for Flow C.
Although $\hat{B}$ is constant, the topology
is different from the Michel's radial-field solution.
The two curves 
$\partial \epsilon / \partial \z =0$
and
$\partial \epsilon / \partial v =0$
meet at large distance,
where an O-type critical point appears.
On the other hand,
the X-type critical point seems to go to the Alfv\'{e}n point
as shown in a closeup view
(the right panel of Figure~\ref{p-fig11}).
The numerical solutions suggest that
both 
${\partial \epsilon }/{\partial v} =0$ and
${\partial \epsilon }/{\partial \z} =0$ 
curves approach the Alfv\'{e}n point tangentially to the $v$-axis.
Thus, the critical solution becomes the vertical line 
coinciding with the $v$-axis as far as $v$ is not so large.
Other solutions are two-valued functions of $\xi$, and are inappropriate.
When $\z=1$ with $v \not= 1$,
the Bernoulli function becomes
\begin{equation}
\left( \frac{\epsilon (1,v) }{mc^2} \right)^2 
= \frac{1}{1-\lambda}
  \left[ 
         1 - \left(\frac{1-\lambda}{\lambda} \right)^2 \sigma^2 v^2 
   \right].
\end{equation}
It follows from this expression that
in the limit $1-\lambda \rightarrow 0$, in other words, 
the centrifugal-driven limit,
the fast energy is $\epsilon_{\rm f} = (1-\lambda)^{-1/2}$,
regardless of $\sigma$.
Figure~\ref{p-fig12} is drawn for $1 - \lambda = 10^{-4}$, and 
therefore $\epsilon_{\rm f} = 100$. 
In this case, the injection point is always very closed to the Alfv\'{e}n point,
as we intend.
The topology of this type is not new but has been
studied by \citet{1991PASJ...43..569T},
and applied in the context of a disc wind around compact stars.

The Lorentz factor of the flow and the azimuthal velocity are
shown in Figure~\ref{p-fig12}.
It can be seen that the flow energy is injected totally as kinetic energy.
Since $\lambda = \Omega \ell / \epsilon \approx 1$, angular momentum is
brought in as well.

In contrast with Flow A and Flow B, 
we cannot determine the value of $\lambda$ and
therefore $\gamma = (1-\lambda)^{-1/2}$ 
since the critical solutions are degenerated.
Determination of $\lambda$ is postponed until
a self-consistent treatment of plasma injection and determination of the poloidal field.

Pair creation in the vicinity of the Y-point is one of the injection process, but
a most efficient way would be magnetic reconnection.
At the top of the closed field region,
a high density corotating plasma with a large Lorentz factor accumulates,
and is likely to be transferred to the open field region via magnetic reconnection.
In this case the value of $\sigma_0 / \gamma $ in (\ref{thickness})
would be very small,
meaning that the toroidal current is enough strong to open the magnetic field,
where note again that the face value of $\sigma_0$ is reduced in Flow C.
The reconnection process will be associated with plasmoid emission
(the dashed small loop in Figure~\ref{p-fig10}).
The injection may be intermittent.
We suppose that the plasmoid formation seen
in PIC simulations corresponds to this injection process.

Since the MHD equation of motion in the cold limit is equivalent to
the equation of motion of individual particles,
it is likely that the motion of plasmoids follows the solution of Flow C.

\section{Pre-acceleration in the closed field region}

Before the injection, 
{ the}  plasma that resides in the closed field region must be accelerated in
the azimuthal direction.
In a steady state, such plasmas migrate from the inner magnetosphere to 
the top of the closed region.
This type of flow can also be described by the field-aligned equation.
Let us call this Flow D.
We again assume $\sigma_0 = 10^4$, and look for a solution with the
expected properties.
The only parameter we can change is $\hat{B}$.
We find that the expected flow appears if $\hat{B}$ increases as $\z \rightarrow 1$.
It is know in the force-free model that the poloidal field strength diverges 
as the Y-point is approached to the light cylinder
\citep{1990SvAL...16...16L,2003ApJ...598..446U,2006MNRAS.368.1055T}.
This increase of magnetic field plays a key role in Flow D.
\begin{figure}
\includegraphics[width=\columnwidth, page=11]{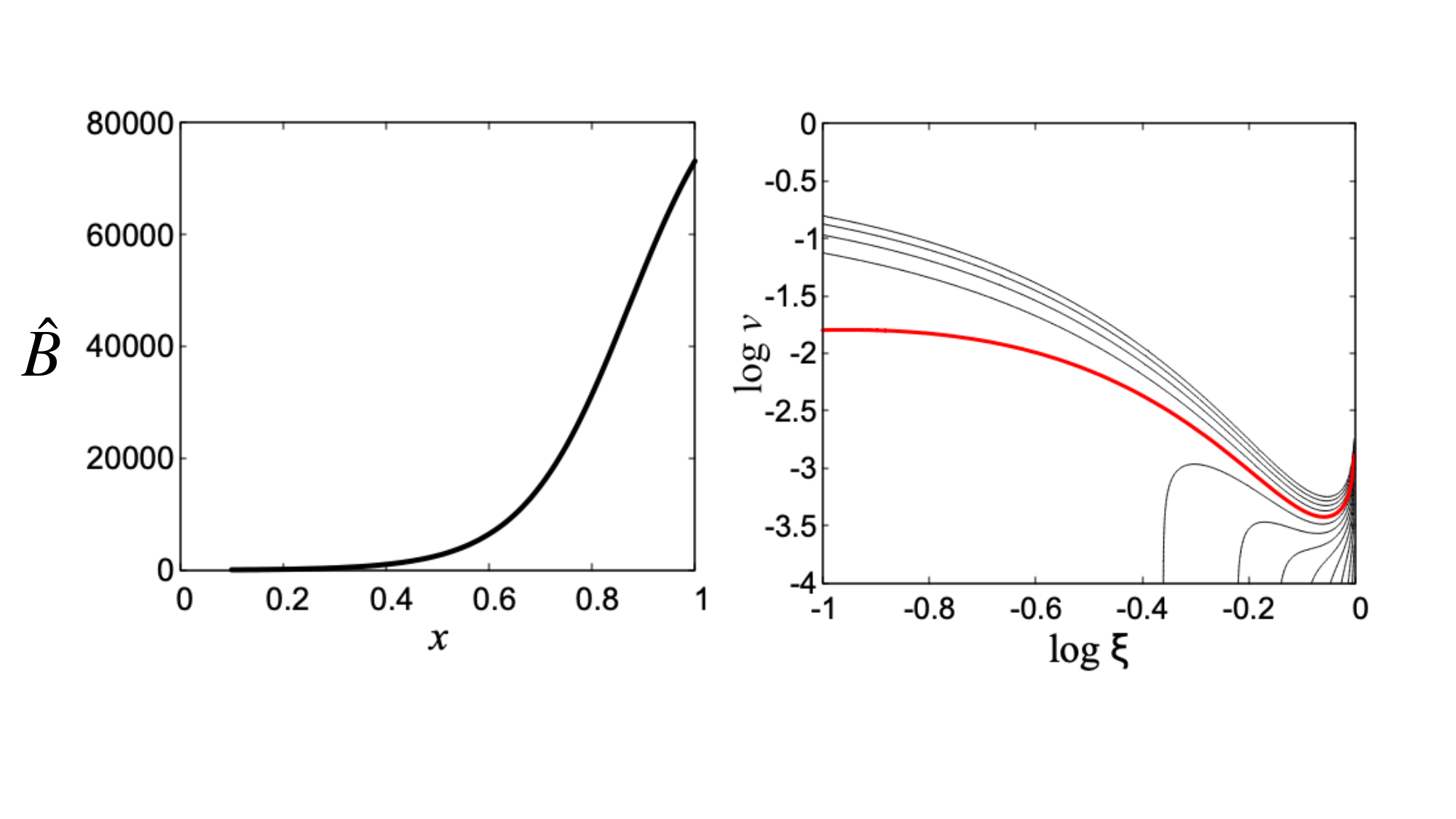}
    \caption{$\hat{B}$ (left) and the family of the solutions (right) of Flow D.}
    \label{p-fig14}
\end{figure}

\begin{figure}
\includegraphics[width=\columnwidth, page=12]{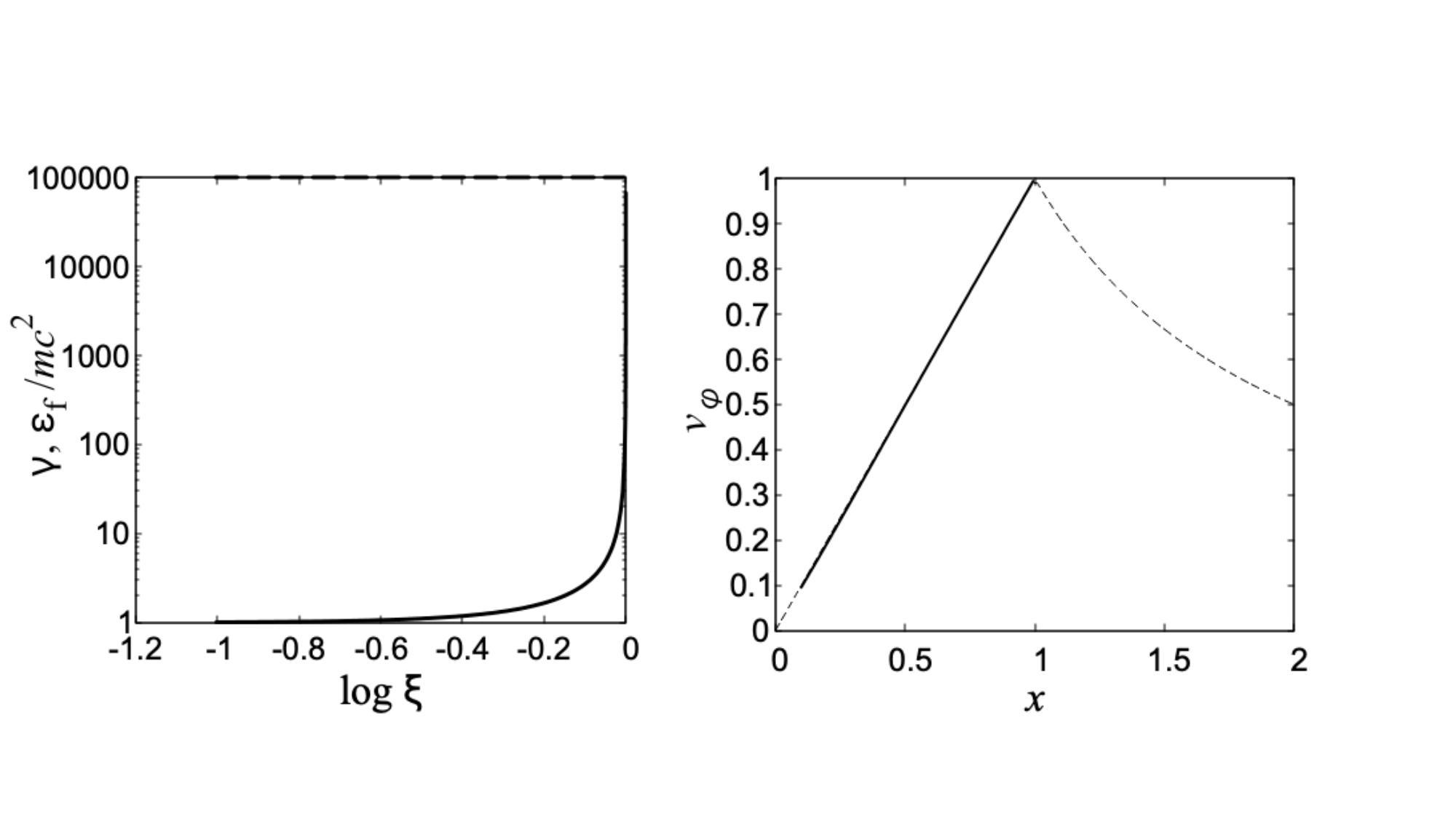}
    \caption{The Lorenz factor (solid line) 
and the energy of the flow (dashed line) as functions of $\log \z$,
and the azimuthal velocity as a function of $x$ for Flow D.}
    \label{p-fig13}
\end{figure}

To give an illustrative solution for Flow D,
we use the same model (\ref{Bhatmodel}) for $\hat{B}$, but
we take a set of parameters, $D=10^5$, $x_d = 0.9$, and $\delta_{\rm d}=0.1$,
mimicking a quick increase of $\hat{B}$ just before the Alfv\'{e}n point.
We denote this model by $\hat{B}_{\rm d}(x)$ as shown in Figure~\ref{p-fig14}.
The right panel shows the family of the solutions for 
$\sigma_0 = 10^4$ and $1- \lambda = 10^{-5}$.
The value of $\lambda$ is chosen so that $\gamma = 1 $ at injection.
As shown in Figure~\ref{p-fig13},
the Lorentz factor grows up to $\approx \hat{B}_{\rm d} (x_{\rm A}) \approx 10^5$
as $\xi \rightarrow 1$, and the azimuthal 
velocity follows the corotation. 
Since $\sigma_0/\gamma \ll 1$, the toroidal current is large enough to 
open the magnetic field.

After some numerical works, we find that
$\epsilon \approx \hat{B}_{\rm d} (x_{\rm A} )$,
$1- \lambda \approx \epsilon^{-1}$,
and 
$\gamma \rightarrow \hat{B}_{\rm d} (x_{\rm A})$,
provided  that the injection point is well within the light cylinder,
the Lorentz factor is $\approx 1$ there, and the azimuthal velocity follows
the corotation throughout the flow.

Although the increase of the poloidal field strength in the force-free model
is due to the imposed condition of
the closed-to-open-field structure via an electromagnetic force balance, 
we find that this increase is related to increase of inertia and in turn
the centrifugal drift current to open the magnetic field.
Therefore, a self-consistent treatment of the field-aligned equation and
the trans-field equation is expected to give the terminal Lorentz factor 
and the saturated field strength at the top of the closed field region.
Furthermore, the transition from Flow D and Flow C would take place, and
therefore magnetic  reconnection must be included in the self-consistent treatment.

\section{Discussion}

Our results seem aligned well with
the recent results of PIC simulations
\citep{2014ApJ...795L..22C, 2015MNRAS.449.2759B, 2023ApJ...943..105H, 2023ApJ...958L...9B, 2022ApJ...939...42H},
in which kinetic energy dominant flows appear around the equatorial
current sheet, and  plasmoids are ejected 
from the top of the last closed field lines.
{
More importantly,
\citet{2022ApJ...939...42H} showed that 
plasmas are accelerated in the azimuthal direction with high Lorentz factors
near the top of the closed field region, and that
the plasmas are released away freely 
by reconnection and plasmoid emission.
These two steps correspond to Flow D and Flow C, respectively.
Comparison between their results and our results would be a future issue.
}

A problem found in the previous sections is that
our model does not predict the thickness of the current sheet
or the Lorentz factor to be reached.
This is because the fast critical point degenerates into the Alfv\'{e}n point 
under the circumstances of equipartition between the magnetic field and the plasma
energy densities ($\sigma \sim 1$).
We need to undertake reconnection processes.
However, before that, we can make an estimate
by assuming that
the thickness is of order of Larmor radius
\citep{2004JGRA..109.2212A,2020ApJ...900...66H}.
We employ (\ref{eqdriftcurrent}), where we set $\Bp^{\rm (out)} \approx \BL$, 
and the density is the value in the Y-point, denoted by $n_{\rm Y}$,
which is distinguished from  the density outside the current sheet, $n$.
With the condition $\Delta \sim (eB/\gamma mc^2)^{-1}$,
where we assume $B \approx \BL$ as a mean value of a strong field in
the closed region and a weak field in the open region,
(\ref{eqdriftcurrent}) yields
\begin{equation}
1 \approx \frac{ 2 {\cal M} \gamma^2 (mc^2)^2 }{(e \BL \RL )^2 }
\frac{\BL}{\BY} \frac{n_{\rm Y}}{n}.
\end{equation}
The continuity implies $(\BL / \BY )(n_{\rm Y}/n) \approx 1$
for the both sides of the Y-point. 
Then we have
\begin{equation}
\gamma \approx 
\frac{\gamma_{\rm max}}{(2 {\cal M})^{1/2}},
\ \mbox{and }
\frac{\Delta}{\RL} \approx \frac{1}{(2 {\cal M})^{1/2}}.
\end{equation}
We also find that the drift velocity is well below $c$ by
a factor of $1/(2 {\cal M})^{1/2}$.
We suggest that the highest energy can be much higher than a simple estimate
of $\gamma_{\rm max}/{\cal M}$.
We may hopefully verify the efficiency factor ${\cal M}^{-1/2}$ by
the future PIC simulations.

A significant difference 
between the present model and the previous PIC simulations 
is the location of the Y-point.
In our previous particle simulations
(see Figure~5 of \citet{2012PASJ...64...43Y}),
we have implied that 
field-aligned potential drop inside the light cylinder makes
a superrotation, and the Y-point move inward.
The PIC simulation by \citet{2022ApJ...939...42H} also shows 
 supercorotation.
In their simulation, there is field-aligned acceleration around the
null surface.
The field-aligned acceleration, superrotation, and subrotation are
interact to each other to establish the proper net loss of the
angular momentum as shown in 
\citet{2021MNRAS.507.1055S}.

Another possible reason for the Y-point within the light cylinder is
that $\lambda$ becomes much smaller than unity.
{ 
This case can happen if the plasma injected into the magnetosphere
has high density and bring a large amount of energy. The introduced
kinetic and thermal energy contribute partly to drive wind.

It is certain that the acceleration in the azimuthal direction takes
place in the closed field region, so that if ${\cal M}$
is not so large, where Larmor radius becomes comparable with the light
radius, then reconnection would start within the light cylinder.
This also can be a reason that the Y-point seems to locate inside 
the light cylinder.}

{
\citet{2024MNRAS.527L.127C} discussed an  increase 
of the poloidal magnetic field just inside the Y-point.
The increase would be caused by the close-open
structure imposed by hand in the force-free model.
We have found that the increase is essential for
the acceleration in the azimuthal acceleration in 
the sub-Alfv\'{e}nic flow.
Although the electromagnetic force balance with the close-open structure
requires the increase of the magnetic field,
the increase results in the particle acceleration, and
the force balance will be achieved with plasma inertia.
In reality, the closed-open structure is maintained 
by the intermittent plasmoid emission.
The system is then sub-steady.}

\section{Conclusions}

We investigate 
the centrifugal acceleration in an axisymmetric pulsar magnetosphere
under the ideal-MHD approximation.
Anticipating the poloidal field structure 
we have solved the field-aligned equations 
for different field lines 
{ to obtain a hit toward } 
a self-consistent view
of the centrifugal wind.
The poloidal field structure
comes into the flow model through $\hat{B} \propto B_{\rm p} \varpi^2$.

Flows on the open field lines running apart form the equatorial current sheet,
 called Flow A,
are Poyinting energy dominant, 
and there is no efficient centrifugal acceleration.
Flow B that comes into the vicinity of a Y-point and goes out 
in a current layer with finite thickness
is characterized by decrease of $\hat{B}$ just beyond the
Alfv\'{e}n point.
Flow B becomes a super fast flow, and the centrifugal acceleration takes place
efficiently. 
However, the centrifugal drift current is not enough to
open the magnetic field lines.
Flow D is a flow coming from an inner part of the magnetosphere 
and reaches the top of the closed field region. 
The key feature of Flow D is 
an increase of the poloidal field strength toward 
the Alfv\'{e}n point, the loop top. 
The azimuthal velocity follows the corotaion, and the plasma
gains a very large Lorentz factor, $\gamma  \sim \hat{B}(x_{\rm A} )$.
The corotating plasma with a large Lorentz factor produced 
in Flow D is expected to be injected as Flow C
via magnetic reconnection.
Flow C is a super fast flow, 
and goes out in the equatorial current layer.
The region around the junction of the two flows 
{ has} a centrifugal drift current large enough to
open field lines.
A jump from Flow D to Flow C require 
magnetic reconnection and plasmoid emission.

Although we have not investigate the reconnection process,
an simple estimate suggests that the Lorentz factor of the centrifugal driven
outflow goes up to $\approx \gamma_{\rm max} {\cal M}^{-1/2}$.
In a global point of view, 
the centrifugal acceleration takes place in a small fraction 
of $\Delta / \RL \approx {\cal M}^{-1/2}$, and
the most of the spin-down power is carried 
by the Poynting energy in the open magnetic flux.
The role of the centrifugal force is to open the magnetic field and
to establish the poloidal current loop.

\begin{figure}
\includegraphics[width=\columnwidth, page=13]{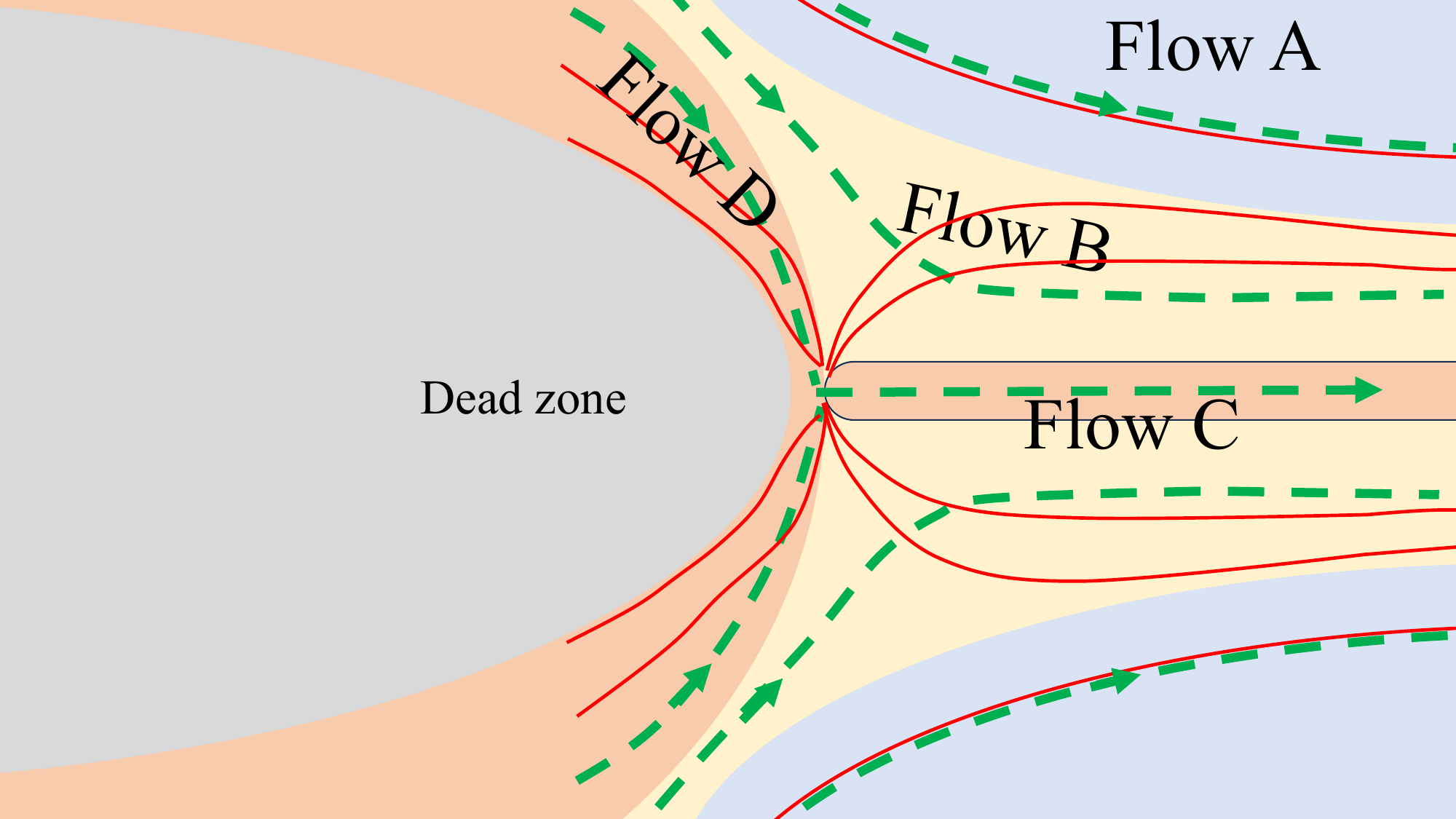}
    \caption{A schematic picture of the vicinity of the Y-point.
The { locations} of Flow A, B, C and D are indicated.
The dashed curves are the streamlines, and the thin curves are the poloidal
current lines.
}
    \label{p-fig15}
\end{figure}

The centrifugal acceleration takes place in Flow B and Flow D just
outside and inside of the Alfv\'{e}n point, respectively.
The poloidal current lines { cross} the stream lines there
as indicated in in Figure~\ref{p-fig15}.
As a result, the current sheets shrinks 
at the Y-point.
This may be a favored poloidal current distribution if 
the regularity of the poloidal field on
the Alfv\'{e}n surface requires an infinitely thin current
as seen in the force-free solution
\citep{1999ApJ...511..351C, 2006MNRAS.368.1055T}.

\begin{acknowledgments}
This work is supported by KAKENHI 
22K03681
(SS, SK), 
21H01078, and 22H01267 (SK).
SS would like to thank Professor Masahiro Hoshino for fruitful discussions
on magnetic reconnection.
\end{acknowledgments}

\bibliography{s}{}
\bibliographystyle{aasjournal}

\end{document}